\documentclass[sigconf]{acmart}
\setcopyright{none}
\AtBeginDocument{%
  }

\copyrightyear{2024}
\acmYear{2024}
\setcopyright{acmlicensed}\acmConference[ICMR '24]{Proceedings of the 2024 International Conference on Multimedia Retrieval}{June 10--14, 2024}{Phuket, Thailand}
\acmBooktitle{Proceedings of the 2024 International Conference on Multimedia Retrieval (ICMR '24), June 10--14, 2024, Phuket, Thailand}
\acmDOI{10.1145/3652583.3658064}
\acmISBN{979-8-4007-0619-6/24/06}

\usepackage{verbatim}
\usepackage{graphicx} 
\usepackage{float} 
\usepackage{subfigure}
\usepackage{multirow}
\usepackage{fancyhdr}
\usepackage{enumitem}
\usepackage{xspace}
\usepackage{balance} 

\newcommand{\ie}{\emph{i.e., }}

\newcommand{\etal}{\emph{et al. }}

\newcommand{\etc}{\emph{etc.\xspace}}

\newcommand{\aka}{\emph{aka. }}

\begin{document}

%\title{Smart Fitting Room: A Generative Approach to Matching-aware Virtual Try-On}

\title{Smart Fitting Room: A One-stop Framework for Matching-aware Virtual Try-on}


\author{Mingzhe Yu}
\affiliation{%
  \institution{Shandong University}
  \country{} 
}
\email{mz_y@mail.sdu.edu.cn}

\author{Yunshan Ma}
\affiliation{%
  \institution{National University of Singapore}
  \country{}
}
\email{yunshan.ma@u.nus.edu}

\author{Lei Wu}
\authornote{Corresponding author.}
\affiliation{%
  \institution{Shandong University}
  \country{}
}
\email{i_lily@sdu.edu.cn}

\author{Kai Cheng}
\affiliation{%
  \institution{Shandong University}
  \country{}
}
\email{k_c@mail.sdu.edu.cn}

\author{Xue Li}
\affiliation{%
  \institution{Jiangnan University}
  \country{}
}
\email{xue.lii754@gmail.com}

\author{Lei Meng}
\affiliation{%
  \institution{Shandong University}
  \country{}
}
\email{lmeng@sdu.edu.cn}

\author{Tat-Seng Chua}
\affiliation{%
  \institution{National University of Singapore}
  \country{}
}
\email{dcscts@nus.edu.sg}

\renewcommand{\shortauthors}{Mingzhe Yu et al.}


\begin{abstract}
The development of virtual try-on has revolutionized online shopping by allowing customers to visualize themselves in various fashion items, thus extending the in-store try-on experience to the cyber space. Although virtual try-on has attracted considerable research initiatives, existing systems only focus on the quality of image generation, overlooking whether the fashion item is a good match to the given person and clothes. Recognizing this gap, we propose to design a one-stop Smart Fitting Room, with the novel formulation of matching-aware virtual try-on. Following this formulation, we design a Hybrid Matching-aware Virtual Try-On Framework (HMaVTON), which combines retrieval-based and generative methods to foster a more personalized virtual try-on experience. This framework integrates a hybrid mix-and-match module and an enhanced virtual try-on module. The former can recommend fashion items available on the platform to boost sales and generate clothes that meets the diverse tastes of consumers. The latter provides high-quality try-on effects, delivering a one-stop shopping service. To validate the effectiveness of our approach, we enlist the expertise of fashion designers for a professional evaluation, assessing the rationality and diversity of the clothes combinations and conducting an evaluation matrix analysis. Our method significantly enhances the practicality of virtual try-on. The code is available at \url{https://github.com/Yzcreator/HMaVTON}.
\end{abstract}



%   \includegraphics[width=1.0\textwidth]{image/front page.pdf}
%   \setlength{\abovecaptionskip}{0.1cm} 
%   \setlength{\belowcaptionskip}{0.cm} 
%   \caption{Mix-and-match and try-on are two essential fashion needs in daily life. Instead of separately modeling, we propose a one-stop system of Smart Fitting Room, in which the user only needs to provide a partially masked image as query, our system will generate an apparel to match with the query and put it on the query image.}
%   \label{fig:frontpage}


\maketitle

\begin{figure}
  \includegraphics[width=0.48\textwidth]{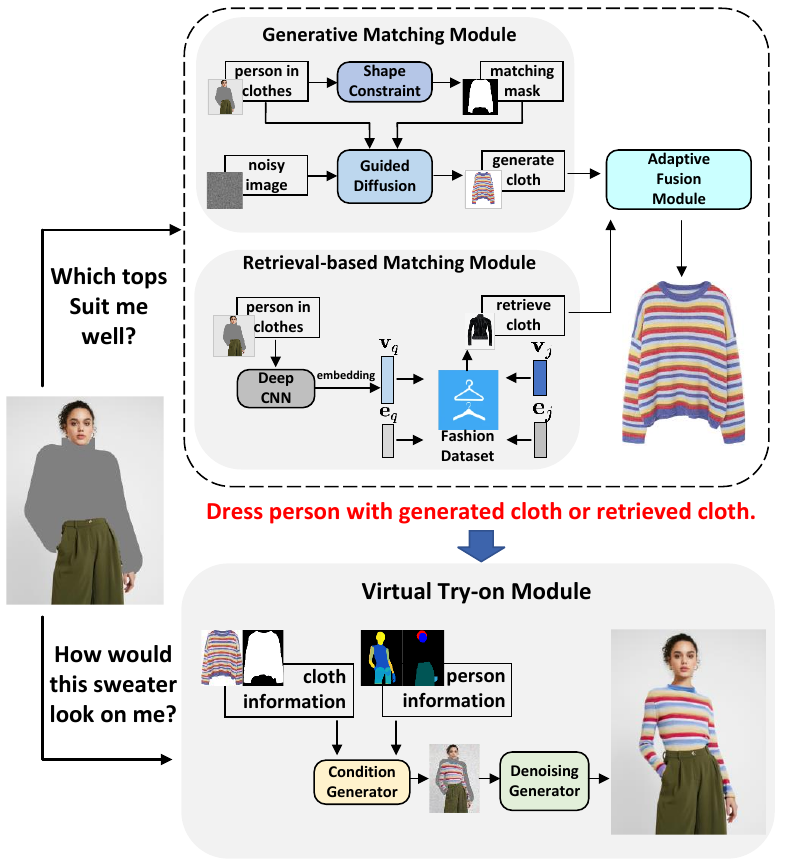}
  \setlength{\abovecaptionskip}{-0.2cm} 
  \setlength{\belowcaptionskip}{-0.7cm} 
  \caption{Mix-and-match and try-on are two essential fashion needs in daily life. We propose a one-stop system of Smart Fitting Room, which will generate or retrieve an apparel to match with the query and put it on the query image.}
  \label{fig:frontpage}
\end{figure}

\section{Introduction}
% First, introduce the current research status in the Try-on domain.
%The task of virtual try-on aims to provide visual images of the wearing effect by using given clothing and human images. 
Virtual try-on facilitates customers in seamlessly envisioning their appearance while perusing diverse fashion items during online shopping. 
%An increasing number of e-commerce platforms are incorporating online try-on as an extension of physical store sales to enhance brand value, reduce the time and cost associated with returns and exchanges, and improve the overall shopping experience. 
It extends the physical world try-on experience to the cyber space, enhancing the shopping experience for customers and thereby mitigating return and exchange rates, while concurrently bolstering sales and profits.
%In recent years, virtual try-on has attracted a lot of attention, and the current methods~\cite{GP-VTON, HR-VTON, ge2021parser, FS-VTON, DCI-VTON, TryOnDiffusion} are capable of achieving promising results in terms of clothing fit and texture retention.
Owing to its considerable value, virtual try-on has attracted growing interest from both academic and industrial community, leading to the emergence of numerous research works~\cite{GP-VTON, HR-VTON, PF-AFN, FS-VTON, DCI-VTON, TryOnDiffusion}.

%Even though the impressing achievements of the current virtual try-on works, there are still some issues. 
Despite significant advancements, the task of virtual try-on has overlooked a crucial aspect within the realm of online shopping, that is mix-and-match. 
Initially, Han \etal define the task virtual try-on~\cite{VITON} as putting on a given fashion item to a partially masked query image. Hence, it only focuses on the quality of image generation regarding human pose, clothes deform and visual patterns, while overlooking a key question: whether the fashion item is a good match to the given clothes and person? 
In essence, this question corresponds to the task of fashion mix-and-match, a classical problem in both conventional fashion research and the recently fast-evolving computation fashion~\cite{FashionRecSurvey-23}. Especially in the community of multimedia and information retrieval, researchers have developed various algorithms, including both retrieval-based~\cite{TransNFCM-19, VBPR, Deepstyle, ACF} and generative methods~\cite{CRAFT, DVBPR, Compatibility}, for fashion mix-and-match. Naturally, mix-and-match and virtual try-on are two essential fashion needs that are highly correlated with each other. Imagine a common shopping scenario: customers would first pick clothes, then try-on, and pick another. Such an interleaved iterative process repetitively plays everyday in online shopping, therefore, it motivates us that why don't we integrate these two separate tasks into a unified one-stop framework?  
%Specifically, current studies fall short of meeting users' styling needs, as many people lack the necessary domain knowledge about fashion mix-and-match. 
%Moreover, the proliferation of various shopping platforms, each offering a diverse range of candidate fashion items, presents a challenge even for experts. 
%They struggle not only with the huge candidate set but also with the lack of interoperability or availability of clothing across platforms. These issues render existing virtual styling solutions imperfect for aligning with users' need for automatic mix-and-match. 

%In light of the aforementioned issues, we aim to develop a one-stop system, Smart Fitting Room. 
Following this motivation, in this paper, we seek to develop a novel system, \ie Smart Fitting Room, for the novel task of matching-aware virtual try-on, as shown in Figure~\ref{fig:frontpage}. 
%The novel one-stop formulation not only makes it more convenient for online shopping, but also xxx. Here I want to hightlight the novel formulation can make it possible to jointly training for both mix-and-match and virtual try-on, but we have not implemented this. So I just do not mention it. It should be noted that here the motivation of combine two tasks into one is weak. which can consecutively fulfill both tasks of mix-and-match and virtual try-on. 
This system provides users with diverse, well-coordinated, and stylish clothing combinations, along with visual images of the virtual try-on effects.
Compared with the conventional virtual try-on systems, our system can put on diverse and well-matched clothes to the given person. At the same time, comparing to the previous mix-and-match methods, our system is capable of demonstrating the synthesized try-on effects to the users, thus improving the consumers' shopping experience and the convert rate of the platforms. 
Despite various benefits, developing such a system is not simply chaining two separate models. Conversely, the distinctive motivation and characteristics of these two tasks pose exceptional challenges to the implementation of this novel system.
An effective mix-and-match model often requires a large-scale dataset with tens of thousands of fashion items. It is not only necessary for training but also crucial for diversified recommendations since a limited number of options can hardly satisfy consumers' various fashion tastes. However, most of the online fashion stores can never access such a large-scale dataset. Moving one step back, even though they are able to procure such large-scale dataset in some way, it is still difficult for them to benefit from the external dataset. This is because existing mix-and-match models retrieve or generate items unrestricted, it is inevitable to recommend some items that are not served by itself or even non-existent, and recommending such items is commercially helpless to the pertinent store. In summary, there are long-standing yet overlooked challenges in developing a one-stop system for smart fitting room.
%Consequently, incorporating a virtual try-on system to online fashion stores is easy and . In contrast, the mix-and-match system usually requires tens of thousands fashion items during training, more impo.  designed for the entire platform with  do not apply or benefit to individual brands or shops. 
%(1) How to design a framework that integrates matching information and virtual try-on data. (2) Designing a matching model that integrates seamlessly with the try-on functionality in the framework is challenging. 
%In this regard, fashion researchers leverage recommendation models that can pick up the top-ranked clothes from a large set of candidates to match with a given fashion item. 
%However, there are two concerns in practice: (1) training such a mix-and-match model often requires large-scale training data of high-quality matched clothing pairs, which are often unaccessible for small platforms; and (2) to assure the matching quality and diversity, recommendation models need a large set of candidate fashion items, which is also unaffordable for individual small platforms.

Addressing the above challenges, we propose Hybrid Matching-aware Virtual Try-On Framework (HMaVTON), encapsulating a hybrid mix-and-match module and an enhanced virtual try-on module. Specifically, we innovatively combine two mix-and-match models, which are separately trained following two distinctive paradigms, \ie one is retrieval-based and the other is generative. Upon two sets of recommended items, we design a simple yet effective adaptive fusion method to ground the generated items to the retrieved items. As a result, we can offer a hybrid list of fashion items, where the grounded items that exist on the platform can help boost sales and profits, while the generated items can satisfy consumers' diverse tastes and improve the shopping experience. To be noted, our fusion method is controllable with a simple threshold, therefore, we can smoothly change the ratio between generated and retrieved items. We integrate this hybrid mix-and-match module to an enhanced virtual try-on module, resulting in the final framework HMaVTON. 
%To address the challenges mentioned above, we propose a generative approach to matching-aware virtual try-on, integrating style information into the generation process. 
%Specifically, we integrate a generative matching module to a virtual try-on model, thus achieving a unified matching-aware virtual try-on approach. 
%Thereby, given a query image with partially masked clothes, as shown in Figure \ref{fig:frontpage}, our approach can generate an apparel to match with the given clothes, and produce virtual try-on image of the character wearing the clothes. 
%To design a matching model that integrates well with try-on, we employ a generation-based matching approach. This approach leverages an external large-scale matching dataset to pre-train the matching model, thus to transfer the matching knowledge into the try-on domain. 
%In terms of model evaluation, since this task lacks quantitative evaluation metrics, 
In terms of the evaluation, since there is no available ground-truth for matching-aware try-on systems, we collaborate with fashion designers and conduct expert-level human evaluation. The experimental results demonstrate that our framework offers one-stop shopping service. In particular, the hybrid mix-and-match module can yield best matching score, and our enhanced virtual try-on module can generate try-on effects of higher quality.
%users are no longer constrained to trying on individual pieces of clothing but can explore a wider range of fashion choices based on their personal style and preferences. 
%This significantly enhances the practicality of virtual try-on and caters to users' broader fashion experience needs.
In summary, \textbf{the primary contributions} of our work are as follows:

\begin{itemize}[leftmargin=0.5cm, itemindent=0cm]
 \item We introduce a novel task of matching-aware virtual try-on, which is the first time to integrate two essential fashion needs, \ie mix-and-match and virtual try-on, into a unified framework.
 %where users are no longer confined to sifting through clothing lists but rather explore a broader range of fashion combinations under our mix-and-match advice.
 \item We present the Hybrid Matching-aware Virtual Try-On Frame-work (HMaVTON), where the hybrid mix-and-match module adaptively fuses both retrieval-based and generative matching results, reaching a controllable balance between user experience and commercial benefits.
 %incorporating external pairing information and designing a generative matching model to replace the retrieval matching model.
 \item We collaborate with fashion designers and conduct an expert-level human evaluation. And the results indicate that our framework can yield best performance.
 %We collaborated with fashion designers researchers, proposing a professional clothing evaluation method. Fashion design experts were invited to participate in the evaluation, and the results indicated that our matching clothes were recognized on styles, colors, and other aspects.
\end{itemize}

\section{Related Work}

\begin{figure*}[h]
  \centering
  \setlength{\abovecaptionskip}{0.1cm} 
  \setlength{\belowcaptionskip}{-0.4cm} 
  \includegraphics[width=1.0\textwidth]{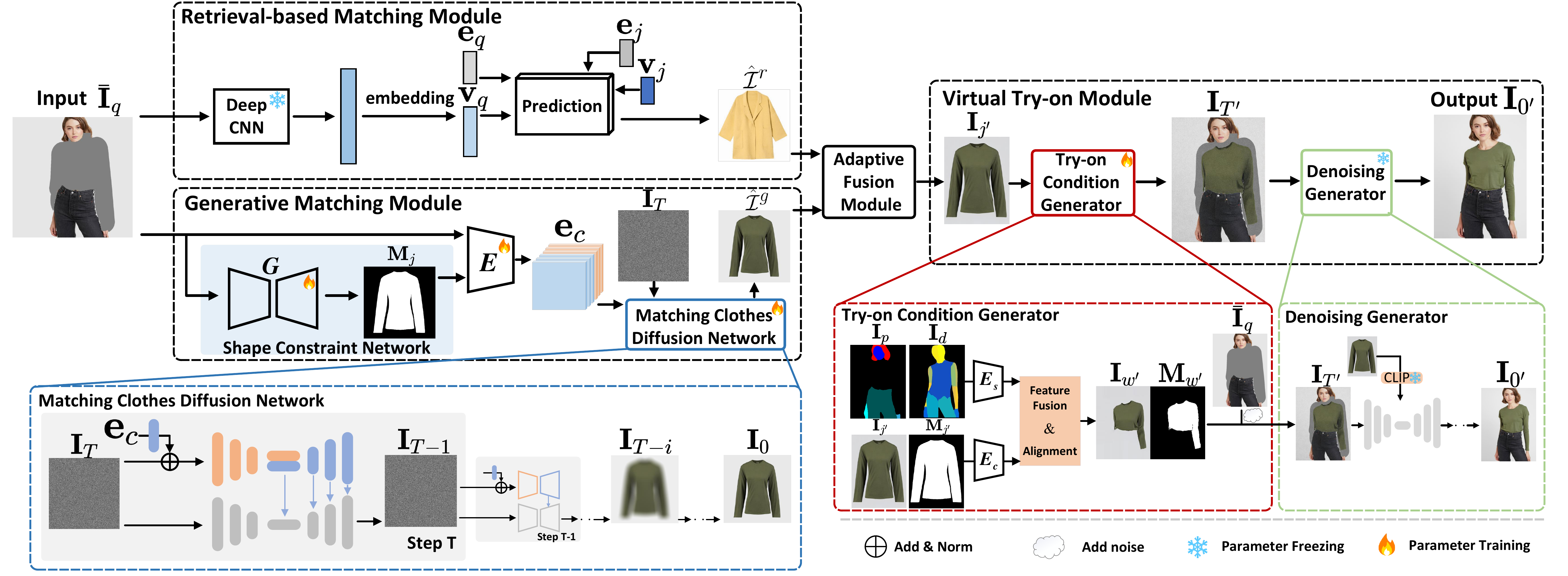}
  \caption{
  A schematic of our one-stop framework. We adaptively fuse retrieval-based and generative matching results using the Adaptive Fusion Module, and provide high-quality try-on effects through the Virtual Try-on Module.
  }
  \label{fig:model}
\end{figure*}
 
\noindent \textbf{Virtual Try-on:} Virtual try-on~\cite{FashionOn, TF-TIS, Fashionmirror} aims to seamlessly transfer clothing onto specific characters. Previous works~\cite{CP-VTON, VTNFP, Zflow, OVNet, GP-VTON} involves generating warped clothes aligned with character, and then generating images of the character wearing the warped clothes. In the clothes warping stage, VITON~\cite{VITON} introduce a process guided by TPS~\cite{tps} for local non-rigid deformation of clothes. Subsequent research~\cite{ClothFlow, FS-VTON, DAFlow, POG} introduce flow-based methods. In the clothes and character synthesis stage, previous approaches~\cite{VITON, PF-AFN, FS-VTON, HR-VTON, GP-VTON} have predominantly relied on GANs, which are susceptible to mode collapse during training. 
%Consequently, GANs face challenges and limitations in the context of virtual try-on tasks, particularly in scenarios requiring detailed try-on effects. Therefore, 
Current research~\cite{DCI-VTON, TryOnDiffusion} are increasingly adopting diffusion-based methods. But users are constrained by the clothes items provided in the dataset, making it challenging to validate the rationality of their outfit choices. 

\noindent \textbf{Fashion Mix-and-Match:} Mix-and-match is a popular task in the domain of computational fashion~\cite{FashionRecSurvey-23}, where the major solutions are based on item-based recommendation models for either pair-wise matching~\cite{TransNFCM-19, AIC, LiaoDM23, DingMBY023, YangDW20} or outfit recommendation~\cite{TOG-23, CLHE}. 
%The fashion industry, influenced by market dynamics and customer preferences, possesses a wide range of clothes catalogs. 
However, the scarcity of user interactions with clothes items complicates the use of standard recommender systems. In previous works, fashion mix-and-match clothes often utilizes visually-perceived collaborative filtering techniques. Approaches like VBPR~\cite{VBPR} extract image category features and augment item vectors with them. 
%DeepStyle~\cite{Deepstyle} uses a User-Item matrix to learn item style, modeling them as a combination of style and category. 
But they are constrained by existing clothes datasets and can only recommend pre-existing clothes pairings. The generative fashion recommendation models~\cite{CRAFT, DVBPR, yang2018recommendation, Compatibility, cGAN, DiFashion} have the capability to create novel fashion mix-and-match clothes, providing more creative and personalized fashion suggestions. Previous works, such as CRAFT~\cite{CRAFT}, generate feature representations for clothes pairings and retrieve the most suitable individual clothes items from the dataset. DVBPR~\cite{DVBPR} generates clothes images based on user preferences but is limited to generating images that are identical in shape to those in the dataset. Despite these efforts, they are still constrained by the dataset and often produce unsatisfactory results. We introduce a novel diffusion model-based fashion recommendation model that not only produces high-quality and diverse clothes items but also generates visual images of users wearing clothes, significantly enhancing the user experience.
\section{Method}
We first present the problem formulation our proposed new task of matching-aware virtual try-on. Thereafter, we describe our model design, consisting of two key modules: 1) the mix-and-match module, which innovatively unifies both retrieval-based and generative methods for versatile fashion matching; and 2) the virtual try-on module, which optimizes the clothes warping effects to enhance the detail quality of the generated images.

\subsection{Problem Formulation}
We first revisit the problem formulations of mix-and-match and virtual try-on, followed by our proposed novel formulation of matching-aware virtual try-on.

\noindent \textbf{Mix-and-Match.} Given a query fashion item $i$ and its associated image $\mathbf{I}_i \in \mathbb{R}^{3 \times H \times W}$, where $H, W$ are the height and width of the image, respectively, the task of mix-and-match aims to offer another fashion item $j$ that matches well to $i$ in terms of functionality, style, color, \etc There are two typical paradigms for mix-and-match: 1) retrieval-based methods, which aim to retrieve the top-k fashion items from a set of candidate items $\mathcal{I}=\{i_1, i_2, \cdots, i_N\}$, where $N$ is the size of the item set and the ground-truth item $j \in \mathcal{I}$; and 2) generative methods, which aim to train a generation model that directly generates an item $j$, \aka the image of item $j$, where $j \in \mathcal{I}$. 

\noindent \textbf{Virtual Try-on.} We have a ground-truth image $\mathbf{I}_q \in \mathbb{R}^{3 \times H \times W}$ that depicts a person wearing a pair of clothes (\ie the top item $j$ and the bottom item $i$). We mask the area of one item (\ie the top item $j$) in $\mathbf{I}_q$ and obtain the partially-masked image $\bar{\mathbf{I}}_q \in \mathbb{R}^{3 \times H \times W}$. Given the partially-masked query image $\bar{\mathbf{I}}_q$ and the image $\mathbf{I}_j$ that is a product image of the masked fashion item $j$, the task of virtual try-on aims to restore the original $\mathbf{I}_q$, \ie trying this product $j$ on the given person. To be noted, the try-on item $\mathbf{I}_j$ is provided as input, while the core of virtual try-on is make the synthesized try-on image convincing by considering the pose, deformation, and visual patterns~\cite{VITON}.  

\noindent \textbf{Matching-aware Virtual Try-on.} We propose a novel formulation to integrate the function of mix-and-match into the virtual try-on task. Specifically, given a partially-masked image $\bar{\mathbf{I}}_q$ (where the top item $j$ is masked and the bottom item $i$ is shown), this module needs to first recommend an item $j'$ that is well matched to the presented item $i$, then put on its corresponding image $\mathbf{I}_{j'}$ to $\bar{\mathbf{I}}_q$, thus to recover a completely-dressed image $\mathbf{I}_q$. To be noted, the product item $j'$ is not necessarily to be same with the originally-masked item $j$, just to make sure that $j'$ is matched well with $i$ conditioned on the person presented in $\bar{\mathbf{I}}_q$. In practice, the item $j'$ can be retrieved from the candidate set $\mathcal{I}$ or generated from scratch.
%Given a clothing-agnostic image $I \in \mathbb{R}^{3 \times H \times W}$ of a person, we propose the Generative Fashion Matching-aware Virtual Try-on Framework(GMVT). 
%\ymz{It fuses generative and retrieval approaches for clothing matching. The retrieval method utilizes the visual information of images to effectively solve the issue of reduced accuracy caused by data sparsity. The generative method extracts features from image $I$ and, through a diffusion process, generate matching cloth $I_0 \in \mathbb{R}^{3 \times H \times W}$ by incorporating external fashion knowledge. And this framework generates a visual representation $I_0^{'}  \in \mathbb{R}^{3 \times h \times w}$ of person wearing the generated clothing.}

\subsection{Hybrid Mix-and-Match}
Conventional approaches to fashion mix-and-match use either retrieval-based or generative methods, each of which has inherent limitations. Specifically, retrieval-based method can only recommend fashion items from an existing candidate pool, therefore, such method would fail if there is no suitable fashion items in stock on the given platform. In contrast, generative methods can address this problem owing to its capability in generating any fashion item without being limited to a given set of items. However, over reliance on generated items would harm the profits of the platform due to two reasons: 1) the existing fashion items will receive less exposure to users, undermining the sales of existing fashion items; and 2) the generated items require further customization and manufacture to convert to physical products, where the additional costs and longer leading time would depress or even lose the potential users. 

To address the above limitations, we propose a novel hybrid mix-and-match method that takes advantage of both retrieval-based and generative methods, while preventing the pertinent drawbacks. Specifically, we first employ two types of methods to offer mix-and-match recommendations separately, then we propose an adaptive fusion strategy to combine both types of matching results.

\subsubsection{Retrieval-based Matching Module}
We follow one of the typical methods VBPR~\cite{VBPR} to build the retrieval-based matching module. Specifically, given the partially-masked query image $\bar{\mathbf{I}}_q$ and its corresponding matched fashion items $j$, we employ the pre-trained deep CNN model, \ie ResNet-50~\cite{resnet}, to extract the visual features and then leverage a linear layer to transform the visual features into a shared representation space, which is formally defined as:
\begin{equation}
\begin{aligned}
  \mathbf{v}_q &= \text{ResNet}(\bar{\mathbf{I}}_q)\mathbf{W}_1, \\
  \mathbf{v}_j &= \text{ResNet}(\mathbf{I}_j)\mathbf{W}_1,
\end{aligned}
\end{equation}
where $\mathbf{v}_q, \mathbf{v}_j \in \mathbb{R}^d$ are visual representations, $\mathbf{W}_1 \in \mathbb{R}^{2048 \times d}$ is the feature transformation matrix of the linear layer, $\text{ResNet}(\cdot)$ represents the ResNet-50 network, and $d$ is the dimensionality of the visual representation. Thereafter, we can calculate the matching score $\hat{x}_{q,j}$ via the following equation:
\begin{equation}~\label{eq:match_score}
  \hat{x}_{q,j} = \alpha + \beta_q + \beta_j + \mathbf{e}_q^{\intercal}\mathbf{e}_j + \mathbf{v}_q^{\intercal}\mathbf{v}_j,
\end{equation}
where $\mathbf{e}_q, \mathbf{e}_j \in \mathbb{R}^d$ are the id embeddings of query image $q$ and $j$, which are randomly initialized to pertain the collaborative filtering (CF) patterns~\cite{VBPR}. $\alpha, \beta_q, \beta_j$ are trainable parameters to model the bias for the global, query $q$, and item $j$, respectively. We use Bayesian Personalized Ranking (BPR)~\cite{bpr} loss to train the model, denoted as:
\begin{equation}
  \mathcal{L}^{\text{BPR}} = \sum_{(q,j,k) \in \mathcal{Q}}{-\text{ln}\sigma(\hat{x}_{q,j}-\hat{x}_{q,k})},
\end{equation}
where $\mathcal{Q}=\{(q,j,k)|x_{q,j}=1, x_{q,k}=0\}$, $x_{q,j}$ is the ground-truth matching relation between query $q$ and $j$, $x_{q,j}=1$ indicates that $(q, j)$ are matched with each other. In contrast, $x_{q,k}=0$ means $(q, k)$ is a an unmatched pair. $\sigma(\cdot)$ represents the sigmoid function. During inference, given a partially-masked query image, we rank all the items in the candidate set using the score function (defined in Equation~\ref{eq:match_score}) and take the top-k ranked items as the matched items, represented as $\mathcal{\hat{I}}^{r}=\{i_n^r\}_{n=1}^{k}$.  

\subsubsection{Generative Matching Module}
The great success of Stable Diffusion~\cite{stable_diffusion} and its following work ControlNet~\cite{controlnet} bring new opportunities for generative mix-and-match. Therefore, we propose a novel generative matching module based on ControlNet. Specifically, we first employ a Shape Constraint Network to generate the mask that depicts the shape of the desired fashion item image, then we use ControlNet the generated the matched image, by taking the generated mask as well as the original partially-masked query image as control conditions. 

\noindent \textbf{Mask Generation via GAN.}
We utilize a GAN~\cite{gans} characterized with U-net~\cite{u-net} to generate the mask of the clothes to be generated, which is called Shape Constraint Network. Specifically, the generator $G$ takes the partially-masked query image $\bar{\mathbf{I}}_q$ as input and generates the mask $\mathbf{M}_j$, which corresponds to the shape of the ground-truth product $j$. To be noted, the mask $\mathbf{M}_j$ is for the product image instead of the masked area in the query image. The discriminator $D$ takes in $\mathbf{M}_j$ and discriminates if it a generated mask or the ground-truth mask. To optimized this network, we take advantage of both conditional GAN~\cite{CGANs} and IoU~\cite{Iou}, thus yielding a combined loss function. Owing to such a combination, empirical results show that it can generate high-quality masks. The loss function is defined as:
\begin{equation}
    \mathcal{L}_{G}=\arg \min _G \max _D \mathcal{L}_{\text{cGAN}}(G, D)+\lambda \mathcal{L}_{\text{IoU}}(G).
\end{equation}

\begin{figure*}
  \centering
  \setlength{\abovecaptionskip}{-0.3cm} 
  \setlength{\belowcaptionskip}{-0.6cm} 
  \includegraphics[width=1.0\textwidth]{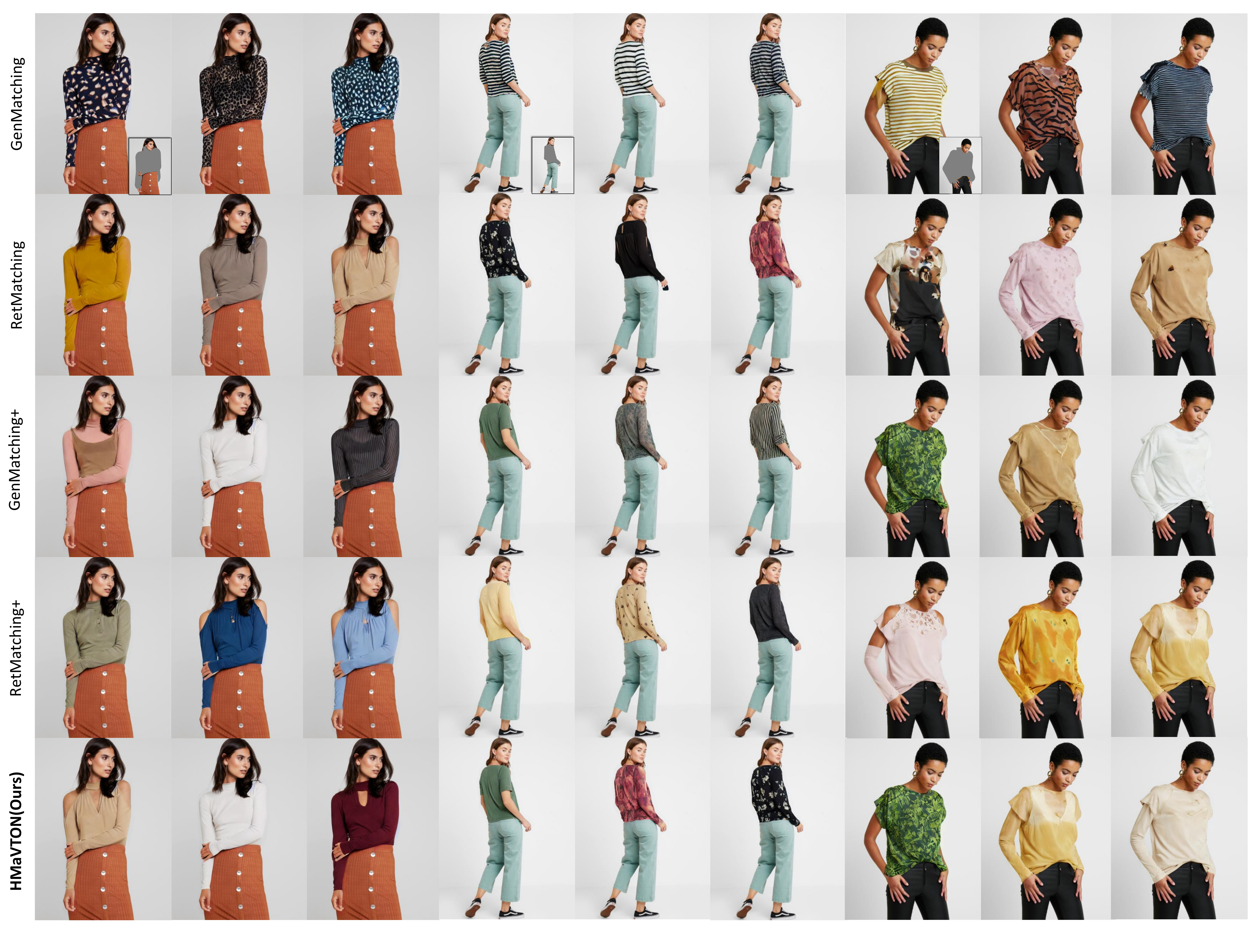}
  \caption{In comparison with baseline, our approach produces a diverse range of clothes styles, such as long sleeves, short sleeves, and straps, with color matching that adheres to the principles of harmony, creating a soothing visual resonance in terms of brightness and saturation.}
  \label{fig:result}
\end{figure*}

\noindent \textbf{Image Generation via ControlNet.}
We adopt ControlNet~\cite{controlnet} to generate the final matching image. 
Concretely, we use encoder network $E$ to convert pixel-space images ( partially-masked query image $\bar{\mathbf{I}}_q$ and the generated mask $\mathbf{M}_j$) into latent images, then we concatenate both representations to form the control input of the ControlNet, formally represented as:

\begin{equation}
    \mathbf{e}_c = \text{Concat}(E(\bar{\mathbf{I}}_q),E(\mathbf{M}_j)),
\end{equation}
where $\mathbf{e}_c \in \mathbb{R}^{256 \times 64 \times 64}$, and $\text{Concat}(\cdot,\cdot)$ is the concatenate operation on the channel dimension. Thereafter, we leverage a typical ControlNet, and the predicted noise $\epsilon_\theta$ is represented as following: 
\begin{equation}
  \epsilon_\theta = \text{F}(\mathbf{x};\Theta) + \text{Z}\left({\text{F}}\left(\mathbf{x} + \text{Z}\left(\mathbf{e}_c; \Theta_{z}^1\right);\Theta_{c}\right);\Theta_{z}^2\right),
\end{equation}
where $\mathbf{x}$ is the feature of noisy image $\mathbf{I}_t$ and $\text{F}(\cdot;\Theta_c)$ is the neural network blocks in the Stable Diffusion model~\cite{stable_diffusion}, where $\Theta_c$ are the trainable parameters cloned from the original stable diffusion model. $\text{Z}(\cdot;\Theta_{z})$ is a 1x1 convolution layer, where $\Theta_{z}^1$ and $\Theta_{z}^2$ are trainable parameters initialized with 0.

Given the text prompt $p$ and the guiding condition vector $\mathbf{e}_c$, the algorithm learns a denoising network to predict the added noise. The learning objective is represented as:
\begin{equation}
\left.\mathcal{L}_{C}=\mathbb{E}_{{\mathbf{I}_0}, t, p, {e_c}, \epsilon \sim \mathcal{N}(0,1)}\left[\| \epsilon-\epsilon_\theta\left({\mathbf{I}_t}, t, p, \mathbf{e}_c\right)\right) \|_2^2\right],
\end{equation}
where time step $t$ descends from $T$ to $1$ during the denoising process, and the input textual prompt $p$ is simply set as "cloth". More sophisticated textual prompts are left for future study. Through the backward denoising process, we finally obtain the generated matching clothes $\mathbf{I}_{j'}$, \ie $\mathbf{I}_{0}$. Through running the generation process multiple rounds, we can also obtain a list of generated images, denoted as $\mathcal{\hat{I}}^{g}=\{i_n^g\}_{n=1}^{k}$.

\subsubsection{Adaptive Fusion Module}
%This mechanism is used to determine whether the retrieved clothes $\mathbf{I}_r$ or the generated clothes $\mathbf{I}_0$ is ultimately recommended to the user.
After obtaining two sets of matching items $\mathcal{\hat{I}}^{r}$ and $\mathcal{\hat{I}}^{g}$, we propose an adaptive fusion module to combine both results for optimal outcome. The basic idea is to ground the generated images back to the retrieved images. Specifically, for every generated image $i_n^g \in \mathcal{\hat{I}}^{g}$, we calculate the distance between $i_n^g$ and all the retrieved images $\mathcal{\hat{I}}^{r}$, where we use CLIP~\cite{CLIP} to extract the image embeddings and cosine similarity as the distance metric. 
%to encode the generated images $\mathbf{I}_{j'}$ and retrieved top-k images $\mathbf{I}_{r_i}$ into a vector $\mathbf{v}_0$ and $\mathbf{v}_{r_i}$, where $i \in (1,5)$, measure image relevance by calculating cosine similarity, and set a threshold. 
We set a threshold $p$, and if the cosine similarity is larger than $p$, we replace the generated image with the corresponding retrieved item. Therefore, the final results include both existing items and generated items, where the ratio can be adjusted by tuning the threshold $p$. 

\subsection{Virtual Try-on Module}
%This module will dress the clothes on the character. 
When we generate the matched fashion item $i_{j'}$, our next step is to put on this generated image to the partially-masked query image $\bar{\mathbf{I}}_q$. We propose a new try-on model, which includes the Try-on Condition Generator and the Denoising Generator. The former generates the warped clothes $\mathbf{I}_{w'}$ and composites them onto the partially-masked query image $\bar{\mathbf{I}}_q$, while the latter applies noise addition and removal to the composited image, resulting in the visualized try-on image $\mathbf{I}_{0'}$.

%Notably, by adjusting the training loss of the model in the Try-on Condition Generator, we optimize the warp quality of the clothes, thereby improving the model's performance in the details of clothes coordination. The model integrates clothes $\mathbf{I}_c$ and $\mathbf{I}_m$ with human body segmentation images $\mathbf{I}_p$ and $\mathbf{I}_d$, creating suitably warped clothes $\hat{\mathbf{I}_{wc}}$ and $\hat{\mathbf{I}_{wm}}$ through feature fusion and alignment. Image $\mathbf{I}$ combined with the warped clothes undergoes a process of noise addition and denoising, which produces the visualized try-on image $\mathbf{I}^{'}_0$.

\subsubsection{Try-On Condition Generator} This module aims to obtain the warped clothes $\mathbf{I}_{w'}$ and corresponding mask $\mathbf{M}_{w'}$. It employs the concept of appearance flow~\cite{HR-VTON} and consists of two encoders $E_c$ and $E_s$. The clothes encoder $E_c$ is used to extract features from clothes $\mathbf{I}_{j'} \in \mathbb{R}^{3 \times h \times w}$ and mask $\mathbf{M}_{j'} \in \mathbb{R}^{h \times w}$. And the segmentation encoder $\text{E}_s$ extracts features from the segmentation map $\mathbf{I}_d\in \mathbb{R}^{h \times w} $ and $\mathbf{I}_p \in \mathbb{R}^{3 \times h \times w}$. The extracted features are fed into the flow pathway of the feature fusion block to generate the appearance flow map $l_i$. We obtain the warped cloth $\mathbf{I}_{w'}$ and its corresponding mask $\mathbf{M}_{w'}$ through feature fusion and alignment. %in the appearance flow network.

\begin{table*}
  \centering 
  \caption{Results of expert-level human evaluation. Our model has achieved the best performance in terms of overall user satisfaction and score, which indicates that the clothes recommended by our model not only meet the users' needs for matching but also offer them more diverse options.}
  \setlength{\abovecaptionskip}{-0.5cm} 
  \setlength{\belowcaptionskip}{-0.8cm} 
  \vspace{-0.2cm}
  \renewcommand{\arraystretch}{0.3}
    \begin{tabular}{c|cccccccc}
    \toprule Model & Index & Weight & Very Satisfied & Satisfied & Average & Dissatisfied & Very Dissatisfied & Score \\
    \midrule \multirow{5}{*}{GenMatching} & D1-Style & $24 \%$ & 0.070 & 0.300 & 0.388 & 0.182 & 0.06 & \multirow{5}{*}{2.777}\\
     &D2-Color & $25 \%$ & 0.052 & 0.176 & 0.282 & 0.346 & 0.144  \\
     &D3-Fabric & $21 \%$ & 0.054 & 0.224 & 0.344 & 0.296 & 0.082 \\
     &D4-Variety& $30 \%$ & 0.048 & 0.118 & 0.346 & 0.292 & 0.196 \\
     &Weighted score& - & 0.056 & 0.198 & \textbf{0.340} & 0.280 & 0.126 \\
     
    \midrule \multirow{5}{*}{RetMatching} & D1-Style & $24 \%$ & 0.116 & 0.41 & 0.346 & 0.110 & 0.018 & \multirow{5}{*}{3.489}\\
     &D2-Color & $25 \%$ & 0.120 & 0.376 & 0.330 & 0.152 & 0.022 \\
     &D3-Fabric & $21 \%$ & 0.086 & 0.430 & 0.380 & 0.098 & 0.006 \\
     &D4-Variety& $30 \%$ & 0.128 & 0.400 & 0.370 & 0.086 & 0.016 \\
     &Weighted score & - & 0.114 & \textbf{0.403} & 0.356 & 0.111 & 0.016 \\

    \midrule \multirow{5}{*}{\begin{tabular}{c}
         GenMatching+
    \end{tabular}} & D1-Style & $24 \%$ & 0.100 & 0.418 & 0.364 & 0.112 & 0.006 & \multirow{5}{*}{3.420}\\
     &D2-Color & $25 \%$ & 0.090 & 0.348 & 0.346 & 0.200 & 0.016 \\
     &D3-Fabric & $21 \%$ & 0.070 & 0.398 & 0.440 & 0.086 & 0.006 \\
     &D4-Variety& $30 \%$ & 0.076 & 0.420 & 0.390 & 0.106 & 0.008 \\
     &Weighted score & - & 0.084 & \textbf{0.397} & 0.383 & 0.127 & 0.009 \\
     
    \midrule \multirow{5}{*}{\begin{tabular}{c}
         RetMatching+\\
    \end{tabular}} & D1-Style & $24 \%$ & 0.122 & 0.456 & 0.314 & 0.090 & 0.018 & \multirow{5}{*}{3.557}\\
     &D2-Color & $25 \%$ & 0.118 & 0.370 & 0.346 & 0.128 & 0.038 \\
     &D3-Fabric & $21 \%$ & 0.128 & 0.460 & 0.322 & 0.078 & 0.012 \\
     &D4-Variety& $30 \%$ & 0.144 & 0.452 & 0.310 & 0.082 & 0.012 \\
     &Weighted score & - & 0.129 & \textbf{0.434} & 0.322 & 0.095 & 0.020 \\

    \midrule \multirow{5}{*}{\begin{tabular}{c}
         \textbf{HMaVTON(Ours)}
    \end{tabular}} & D1-Style & $24 \%$ & 0.120 & 0.476 & 0.302 & 0.092 & 0.010 & \multirow{5}{*}{\textbf{3.578}}\\
     &D2-Color & $25 \%$ & 0.114 & 0.382 & 0.35 & 0.134 & 0.02 \\
     &D3-Fabric & $21 \%$ & 0.132 & 0.472 & 0.314 & 0.074 & 0.008 \\
     &D4-Variety& $30 \%$ & 0.13 & 0.464 & 0.312 & 0.086 & 0.008 \\
     &Weighted score & - & 0.124 & \textbf{0.448} & 0.32 & 0.097 & 0.012 \\

    \bottomrule
    \end{tabular}
  \renewcommand{\arraystretch}{1}

  \label{tab:userstudy}
  \vspace{-0.3cm}
\end{table*}

During the model training stage, the Try-on Condition Generator use $\mathcal{L}_{TV}$ to enhance the smoothness of the appearance flow:
\vspace{-0.2cm}
\begin{equation}
    \mathcal {L}_{TV} = ||\nabla {l_4}||_1.
    \vspace{-0.2cm}
\end{equation}
And L1 loss and perceptual loss are used to encourage the network to deform clothes to fit the pose of the person:

\vspace{-0.5cm}
\begin{equation}
  \vspace{-0.5cm}
  \mathcal{L}_{L 1}=\sum_{i=0}^3 w_i \cdot\left\|{\rm D}\left(\mathbf{M}_{j'}, {l_i}\right)-\mathbf{M}_{w'}\right\|_1+\left\|\mathbf{\hat{M}}_{w'}-\mathbf{M}_{w'}\right\|_1,
  \vspace{-0.2cm}
\end{equation}

\begin{equation}
    \mathcal{L}_{V G G}=\sum_{i=0}^3 w_i \cdot \phi\left({\rm D}\left(\mathbf{I}_{j'}, {l_i}\right), \mathbf{I}_{w'}\right)+\phi\left(\mathbf{\hat{I}}_{w'}, \mathbf{I}_{w'}\right),
    \vspace{-0.1cm}
\end{equation}
where ${\rm D}(\cdot , \cdot)$ is defined as feature alignment, the operation involves removing non-overlapping regions. 
And $w_i$  represents the weight of relative importance.

In summary, Try-on Condition Generator is trained using the following loss function, where $\lambda_{L1}$, and $\lambda_{TV}$ represent hyperparameters controlling the relative importance between different losses:
\vspace{-0.2cm}
\begin{equation}
    \mathcal{L}_{CG}=\lambda_{L 1} \mathcal{L}_{L 1}+\mathcal{L}_{V G G}+\lambda_{T V} \mathcal{L}_{T V}.
\end{equation}
The warped clothes $\mathbf{I}_{w'}$ and corresponding mask $\mathbf{M}_{w'}$ are 
combined with the partially-masked query image $\bar{\mathbf{I}}_q$, and randomly add $t$ steps of Gaussian noise to obtain $\mathbf{I}_T^{'}$. 

\vspace{-0.2cm}
\subsubsection{Denoising Generator}
The noisy image $\mathbf{I}_T^{'}$ is denoised under the guidance of external condition $\mathbf{I}_{j'}$. The diffusion model loads parameters~\cite{DCI-VTON} as initial parameters. We encode clothes $\mathbf{I}_{j'}$ using the CLIP~\cite{CLIP} image encoder to obtain the conditional input. This condition is injected into the network layers through cross-attention. Since the noisy image $I_T^{'}$ contains the original information of the clothes, the module achieves a more accurate representation of the wearing effect during the reverse diffusion process.

\section{Experiments}

\subsection{Experimental Settings}

\subsubsection{Datasets}
We employ two datasets, \ie POG~\cite{POG} and VITON-HD~\cite{Viton-hd}, where POG is used as an external dataset for the task of mix-and-match and VITON-HD is directly used for the evaluation of virtual try-on and the overall framework. 
%We restrict the frequency of occurrence of each fashion item to be between 5 and 100 times and filter the POG outfit dataset. 
We perform n-core filtering on the POG dataset, where we keep only the fashion items that occur between 5 and 100 times, resulting in 119,978 top-bottom pairs, 14,064 tops, and 8,124 bottoms.
%This process yields a total of 119,978 top-bottom pairs, encompassing 14,064 top items and 8,124 bottom items. 
The VITON-HD dataset comprises 13,679 image pairs, consisting of frontal-view images of women and upper body images.
To mitigate the domain gap between these two datasets during mix-and-match modeling, we perform an additional pre-processing the product images in the POG.
%In the process of introducing matched information, the structural differences between the large-scale matching dataset POG and the try-on dataset VITON-HD's clothes pose a challenge. 
Specifically, for the images of bottom clothes in POG dataset, we adjust their size and move them to the lower part of the overall image. Hence, the bottom images in POG would be better aligned with the partially-masked query images in the VITON-HD dataset. 
%This allows the hybrid mix-and-match model to better learn the structure and features of the paired data.
This allows the hybrid mix-and-match model to focus more on the matching relationship instead of the distribution gap regarding image size and position.

\begin{figure*}
  \centering
  \setlength{\abovecaptionskip}{-0.1cm} 

  \includegraphics[width=0.9\textwidth]{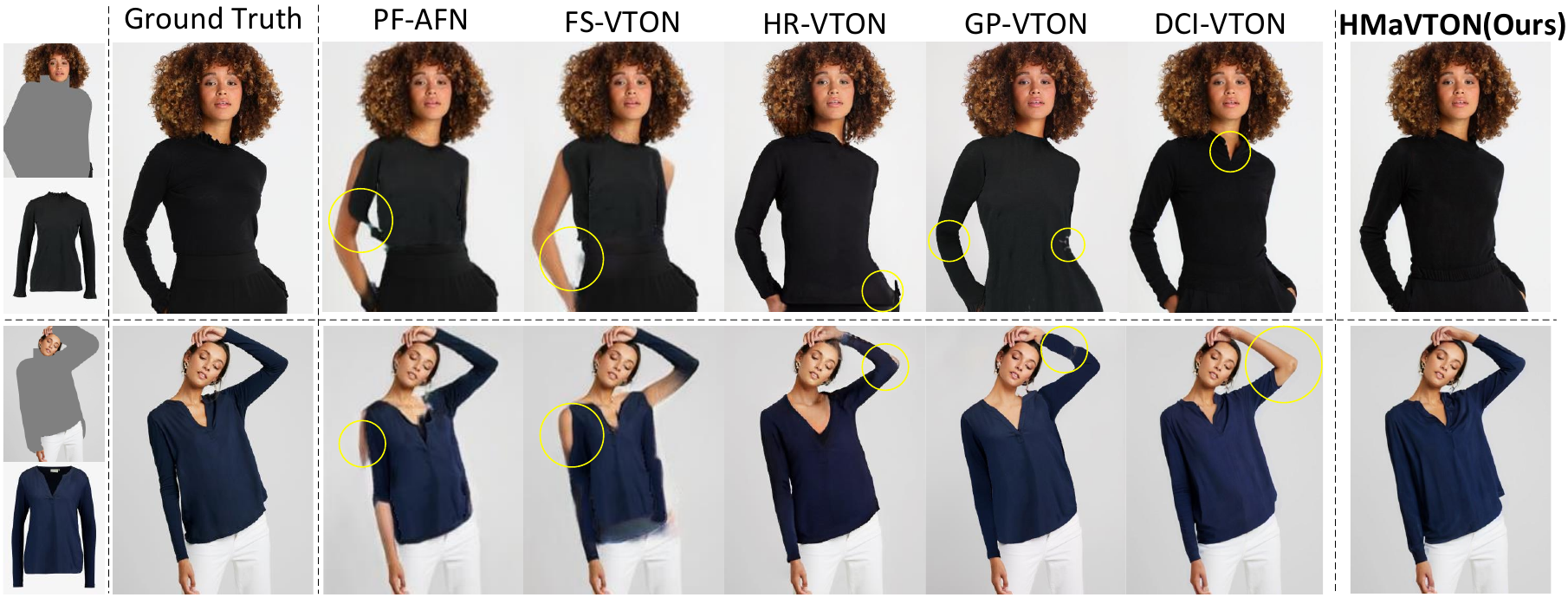}
  \caption{Compared to other try-on models, our method exhibits superior performance in terms of clothes integrity restoration.}
  \label{fig:result_tryon}
  \vspace{-0.3cm}
\end{figure*}

\vspace{-0.2cm}
\subsubsection{Evaluation Metrics} Our task aims to recommend a fashion item and try it on the body. Since the recommended items are not necessary to be identical to the item provided in dataset, how to evaluate the fashion matching and virtual try-on effects is challenging, %This poses great challenge to existing quantitative evaluation metrics.
especially for quantitative evaluation.
To conduct a professional and faithful evaluation, we employ an expert-level human evaluation approach. 
Specifically, we invite a fashion researcher who is majoring in fashion design to propose a list of evaluation protocol and make a questionnaire. Thereafter, we make survey over ten fashion designers and obtain the final quantitative results. 
%Quantitative results are obtained through a weight analysis matrix. 

The evaluation mainly consists of two parts: the rationality of clothes combinations and the diversity of clothes pairing effects. The former is typically analyzed in terms of clothes styles, colors, and fabrics. Specifically, the coordination of clothes styles should adhere to common pairing rules. Color coordination emphasizes the reasonable use and combination of hues. Fabric combinations should align with the season, in which the entire outfit is worn. The diversity primarily focuses on the mutual distinctions in the generated matching combinations, aiming to produce as many different coordination effects as possible.

To evaluate the virtual try-on performance, we employ the following quantitative analysis metrics: SSIM~\cite{SSIM}, FID~\cite{FID}, KID~\cite{KID}, IS~\cite{IS}, LPIPS~\cite{LPIPS}, and PSNR~\cite{PSNR}. Moreover, we use YOLOScore~\cite{yolo} and IS to justify the effectiveness of the SCN.

\begin{table}
  \centering
  \caption{Comparing to other virtual try-on methods, our model achieves the best performance across various metrics.}
  \vspace{-0.3cm} 
  \scalebox{0.8}{
  \begin{tabular}{c|cccccc}
    \toprule
    Method  & SSIM $\uparrow$ & FID $\downarrow$ & KID $\downarrow$ & IS $\uparrow$ & LPIPS $\downarrow$ & PSNR $\uparrow$\\
    \midrule
    PF-AFN & 0.8324 & 35.1156 & 3.0364 & 3.5732 & 0.1026 & 21.3806\\
    FS-VTON &  0.8335 & 35.1156 & 3.1431 & 3.5295 & 0.0913 & 21.9229\\
    HR-VITON &   0.8816 & 9.6404 & 0.2096 & 3.5305 & 0.0847 & 21.9228\\
    GP-VTON &  0.8874 & 9.2379 & 0.1683 & 3.5803 & 0.0713 & 23.0265\\
    DCI-VTON &   0.8952 & 8.3760 & 0.0292 & 3.6785 & 0.0698 & 23.9304\\
    \textbf{HMaVTON(Ours)} &   \textbf{0.8996} & \textbf{8.3349} & \textbf{0.0284} & \textbf{3.6787} & \textbf{0.0671} & \textbf{23.9782}\\
    \bottomrule

  \end{tabular}
  }
  \label{tab:tryon}
  \vspace{-0.6cm}
\end{table}

\vspace{-0.2cm}
\subsubsection{Compared Methods} \label{subsec:compared_methods}
Since the matching-aware virtual try-on task is newly proposed, there is no baseline or SOTA method for this task. To validate the effectiveness of our coordination, we design several baselines, which differ in various coordination modules. Specifically, we categorize the matching models into generation-based models and retrieval-based models. Additionally, to validate the effectiveness of our virtual try-on module, we conduct a comparative analysis of results with the latest try-on models.

\noindent \textbf{Generative Models.} The generation-based models create new clothes combination suggestions based on input images of users. The recommend item  $j'$ is generated from scratch by GMM. We define GenMatching is the baseline that does not incorporate external matching knowledge, while GenMatching+ understands.

\noindent \textbf{Retrieval-based Models.} The retrieval-based model incorporates visual information into recommendation system, returning the top k reasonable clothing combinations from the dataset. The recommend item $j'$ is retrieved from the candidate set $\mathcal{I}$ by RMM. Similarly, we set the baselines for using external matching dataset (POG) RetMatching+ and without introducing it as RetMatching (that only uses the image pairs in the VITON-HD dataset).

\noindent \textbf{Visual Try-on Model.} We also compare the try-on performance between our method and existing try-on models, including PF-AFN~\cite{PF-AFN}, FS-VTON~\cite{FS-VTON}, HR-VITON~\cite{HR-VTON}, GP-VTON~\cite{GP-VTON}, and DCI-VTON~\cite{DCI-VTON}.

\subsubsection{Implementation Details}

% We design a method to introduce large-scale matching information in the try-on domain, addressing the issue of diminished model performance brought about by the scarcity of paired clothing items in try-on datasets. Specifically, we filter the clothing from the POG dataset~\cite{POG} and construct a large-scale matching clothing dataset. The Retrieval-based Matching Module and Generative Matching Module are trained on this dataset to learn ample matching knowledge. Subsequently, the model undergoes fine-tuning on the try-on dataset VITON~\cite{VITON}, incorporating the learned matching information into the try-on domain. This enables the model to better capture the interrelationships and underlying patterns between clothing items, leading to an improved understanding of the ways in which clothing can be matched and the trends in fashion.

Our model is trained on four NVIDIA RTX 3090 GPUs with the Pytorch framework.
The Retrieval-based Matching Module and the Generative Matching Module are both trained on POG dataset and then further fine-tuned on the VITON-HD dataset. The Retrieval-based Matching Module is trained for 100 iterations and fine-tuned for 80 iterations. The Generative Matching Module is trained for 40 epochs and fine-tuned for 50 epochs.

% The execution of our training protocol on the curated POG dataset, spanned across 10 days for a total of 40 epochs. Subsequently, the model underwent fine-tuning on the VITON-HD dataset for 50 epochs, a process that was completed within 2 days. The learning rate was set to 1e-5, and every 300 images, DDIM~\cite{ddim} was used for sampling with 100 sampling steps. 

% The GenMatching Model, in the baseline, was exclusively trained on the VITON-HD dataset, and the training process took 2 days for a total of 60 epochs. The RetMatching+ Model was initially trained on the POG dataset and then further fine-tuned on the VITON-HD dataset, while the RetMatching Model only learned outfit information on the VITON-HD dataset. For the try-on models, we conducted testing on 2032 image pairs using pre-trained models provided by the authors.

\vspace{-0.2cm}
\subsection{Quantitative Results}
Table \ref{tab:userstudy} shows the results of our method and the matching model in human evaluation. We invite ten professional fashion designers aged 18-30 to conduct expert-level human evaluations. 
We use a five-level scoring protocol to score 50 sets of results generated by 4 different models regarding to four aspects of D1-D4 (style, color, fabric, variety).
%The scoring contents are five preference levels: very satisfied, satisfied, average, dissatisfied, and very dissatisfied. 
The scores are discretized to five levels: very satisfied, satisfied, average, dissatisfied, and very dissatisfied, corresponding to 5 to 1 point, respectively. 
The survey is conducted through both online and offline manner. Finally, we collect valid scoring results from 10 designers for analysis. 

\begin{figure}
  \centering
  \includegraphics[width=0.4\textwidth]{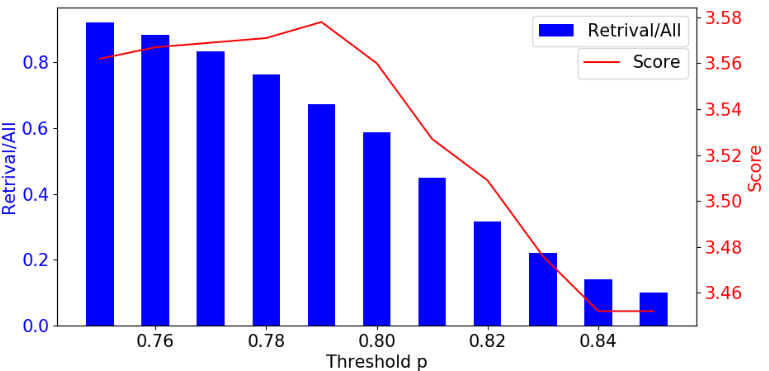}
  \setlength{\abovecaptionskip}{-0.0cm}
  \caption{We balance the retrieval and generation of outfits by controlling the threshold to regulate the proportion of retrieved clothes in all the recommended clothes.}
  \vspace{-0.5cm}
  \label{fig:threshold}
\end{figure}

% \begin{equation}
% \small{
%     \mathrm{R}=\left(\begin{array}{lllll}
% 0.070 & 0.300 & 0.388 & 0.182 & 0.060 \\
% 0.052 & 0.176 & 0.282 & 0.346 & 0.144 \\
% 0.054 & 0.224 & 0.344 & 0.296 & 0.082 \\
% 0.048 & 0.118 & 0.346 & 0.292 & 0.196
% \end{array}\right).
% }
% \end{equation}

We construct the rating matrix $\textbf{R}$ based on the average values obtained from the questionnaire. We calculate the weighted score $\textbf{B} = \textbf{A}\textbf{R}$, where the weight matrix $ \textbf{A} = [0.24, 0.25, 0.21, 0.30] $. The weight matrix $\textbf{A}$ is determined based on the expert's domain knowledge, where five experts first go through the generated results and then decide the weights of the four evaluation metrics D1-D4. The specific weights can be found in Supplementary Material \ref{sec:userstudy}. We calculate the score based on the 5-point Likert scale ratings. Table \ref{tab:userstudy} shows the results. 
%, with a focus on the area with the highest scores in the "Average" row.
Importantly, the scores in the "Weighted Score" row are the final representative results, the best value are highlighted in bold. 
Through the comparison of samples from the four groups, we can see that our model achieves the best performance in overall satisfaction and scores. By incorporating external matching dataset, it can effectively enhance the mix-and-match performance.

\begin{figure}
  \centering
    \setlength{\abovecaptionskip}{0.1cm} 
  \setlength{\belowcaptionskip}{-0.3cm} 
  \includegraphics[width=0.5\textwidth]{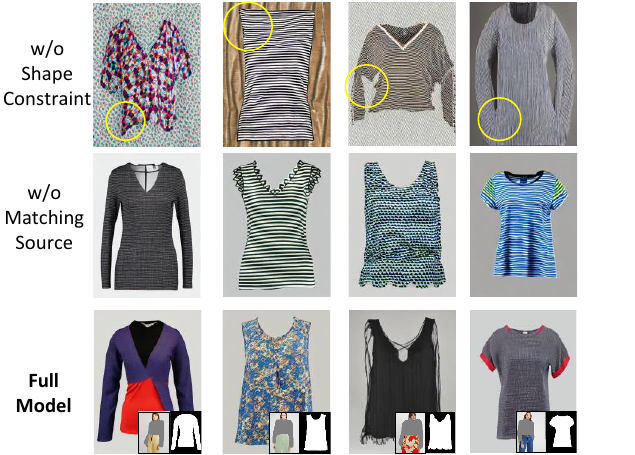} 
  \caption{
  %The validation shows that the Full Model demonstrates better results in terms of image integrity and matching coherence.
  Qualitative ablation study of shape constraint network and matching source.
  }
  \vspace{-0.2in}
  \label{fig:result_ablation}
\end{figure}

Table \ref{tab:tryon} shows the comparison between our model and try-on methods in terms of clothes fitting effects. Notably, the enhancements in the SSIM, FID, and KID metrics show significant improvements, indicating that our generated images closely resemble real images in terms of brightness, contrast, and structure. In the case of the IS metric, there is little variation among different methods, suggesting that the Try-on results are controllable and can genuinely reflect the effect of clothes on the human body, indicating good model stability. Additionally, our model also exhibits slight improvements in the LPIPS and PSNR metrics, indicating an overall perceptual advantage in our images.

Figure \ref{fig:threshold} demonstrates the study of the controllable adaptive fusion module, where the threshold $p$ is set to smoothly
control the 
proportion between generated items and retrieved items. As the threshold increases, the proportion of retrieved images gradually decreases, while the proportion of generated images rises. Notably, the score first increases and then decreases, indicating the existence of an equilibrium point. At this point, the recommended clothes not only exhibit high image quality but also possess a reasonable matching effect while satisfying the users' demands for diversity.

\vspace{-0.2cm}
\subsection{Qualitative Results}
From the left three columns of Figure \ref{fig:result}, we can see that the images generated by our model exhibit better visual results. Not only is there a greater variety of clothes styles, but the fabrics in clothes are mostly soft and skin-friendly. The full-body back view images in the middle three columns demonstrate that our generation module can provide more options for clothes matching. For example, the generated short sleeves may not exist in the clothes database or may not have been recommended, but they still look quite good when worn, which can effectively help users expand their own style. In the right three columns, we can notice that the generated clothes can be mapped back to the retrieved items, and those with distinctive characteristics and novel styles can also be recommended.

Figure \ref{fig:result_tryon} shows that our model maintains better clothes details and offers a richer texture effect in the generated images. In the first row of images, PF-AFN~\cite{PF-AFN} and FS-VTON~\cite{FS-VTON} fail to preserve the overall clothes integrity, HR-VITON~\cite{HR-VTON} exhibits a noticeable protrusion on the right side of the garment's hem, GP-VTON~\cite{GP-VTON} lacks texture on the left joints and contains blurry regions on the right side, while DCI-VTON's~\cite{DCI-VTON} clothes neckline is different from the source attire. In the second row of images, the clothes generated by PF-AFN, FS-VTON, and DCI-VTON all exhibit significant differences from the original attire, HR-VITON and GP-VTON have clothes sleeves that do not conform to body, whereas our generated images perform well in terms of texture and overall appearance.

\begin{table}
  \centering
  \caption{Metrics evaluate the generated results from shape and quality, indicating that the Full Model performs the best.}
  \vspace{-0.3cm} 
  \setlength{\tabcolsep}{4.0mm}{
  \scalebox{1.0}{
  \begin{tabular}{c|cc}
    \toprule
    Method &  YOLOScore$\downarrow$ & IS$\uparrow$\\
    \midrule
    w/o Shape Constraint & 0.1830 & 1.5136\\
    w/o Matching Source & 0.1524 & 2.6420 \\
    \textbf{Full Model} & \textbf{0.1417} & \textbf{2.9149}\\
    \bottomrule
  \end{tabular}
  }
}
  \vspace{-0.5cm} 
  \label{tab:ablation}
\end{table}

\vspace{-0.2cm}
\subsection{Ablation Study}
When setting the baseline in Section~\ref{subsec:compared_methods}, we have validated the effectiveness of incorporating external matching information. Next, we verify the impact of Shape Constraint Network and Matching Source in Generative Matching Module.

\noindent \textbf{w/o Shape Constraint.} Regarding the Generative Matching Module, we design ablated models to validate the effectiveness of the Shape Constraint Network within it. As shown in the first row of Figure \ref{fig:result_ablation}, the absence of shape constraints leads to incomplete garments and boundary blending in the generated clothes.

\noindent \textbf{w/o Matching Source.} To verify the matching capability of our model, we remove the input of human image $I$. From the second row in Figure \ref{fig:result_ablation}, it can be observed that without the inclusion of matching source information, the generated clothes exhibit a monotonous texture and do not match well with bottoms.

Table \ref{tab:ablation} indicates that under shape constraints, our model presents higher image quality and is more in line with aesthetic design. As shown in Figure \ref{fig:result_ablation}, the issues such as incomplete clothes generation are less likely to occur. In terms of the effectiveness of model coordination validation, even without the input of person images, the model can still generate clothes. However, there are cases of monotonous textures and colors, suggesting that the model generates clothes without a clear understanding of the desired style. In contrast, our method not only meets the matching requirements but also ensures high-quality clothes generation.
\vspace{-0.2cm}
\section{Conclusion and Future Work}

We presented a comprehensive generative fashion mix-and-match clothes framework, allowing users to try different clothes combinations under our styling advice. To assess the matching capability of our model, we collaborated with fashion design researchers to design evaluation metrics and invited ten fashion design experts to assess the model. The results indicated that our hybrid mix-and-match approach is well received by the majority of designers in terms of style combinations, color coordination, and other aspects.

In future work, we will enhance the Hybrid Matching-aware Virtual Try-On Framework (HMaVTON) to meet coordination needs of lower garments, shoes, and other clothes items. Simultaneously, we will extend the external matching dataset, considering not only clothes combinations but also incorporating users' profile information. And we will further explore the optimal combination of retrieval-based and generative methods. Furthermore, we will integrate various fashion domain knowledge, such as styles and aesthetics, into our system to enhance the overall system.

\section*{acknowledgement}
This research/project is supported by NExT Research Center.

\newpage
\bibliographystyle{ACM-Reference-Format}
\balance
\bibliography{main}

\newpage
\appendix
\clearpage
\setcounter{page}{1}

\section*{\centering Supplementary Material}

\renewcommand\thesection{\Alph{section}}
\setcounter{section}{0}

\textbf{Overview:} In section \ref{sec:implementation}, we discuss the implementation details of our method and baselines. In the following section \ref{sec:exper},
we present more experiments to investigate the rationality and effectiveness of some detailed model design. Thereby, in section \ref{sec:userstudy}, we introduce user study details and analyze the evaluation data. Finally, we illustrate more generation results in Section \ref{sec:results}.

\section{Implementation Details} \label{sec:implementation}

We propose a new task of matching-aware virtual try-on, for which there currently exist no baselines or SOTA methods. There are some works separately designed for the task of either mix-and-match or virtual try-on, while none of previous works can solve these two problems simultaneously. Hence, we adapt some intuitive models and compose them into several baselines for comparison. These baselines differ with regard to using different mix-and-match modules. We detail the implementation of our method and baselines as follows.

\textbf{HMaVTON:} 
We train the Shape Constraint Network (SCN) with the batch size of 16 and learning rate of 0.0002. The generated shape-constrained image and the original image are concatenated after being encoded into embeddings by a pre-trained VQ-GAN Encoder. This concatenated embedding serves as the external guidance condition for the Matching Clothes Diffusion Network (MCDN) through zero-convolution. The encoder of MCDN takes the parameters of the stable diffusion for initialization, while the parameters of the zero-convolution is initialized with 0. During the training process, we fix the parameters of the stable diffusion and tune the parameters of the encoder, middle block, and zero-convolution. During the sampling process, we employ DDIM for sampling and set the sampling time step as 100. The Virtual Try-on Module utilizes the pre-trained model for clothing deformation and denoising.

\textbf{Compared Methods:} 
We curate several baselines to compare with our method, thus to evaluate the model's mix-and-match capability. We categorize these baselines into Generation-based Model and Retrieval-based Model. GenMatching belongs to the Generation-based model, of which the training batch size is 4, learning rate is 1e-5, and the number of training epochs is 60. We train our model on four NVIDIA 3090 GPUs for the VITON-HD dataset, which takes about two days. RetMatching and RetMatching+ belong to the Retrieval-based Model. The difference between them is that RetMatching+ utilizes the large-scale external mix-and-match dataset of POG while RetMatching only uses the clothes matching pairs in the VITON-HD dataset. All these baseline methods use the Virtual Try-on Module to generate the final images, followed by the same expert evaluation. In order to validate the effectiveness of our Virtual Try-on module, we compare our method with the latest Try-on models. For the baselines, we directly load the pre-trained models released by the authors and test them on the VITON-HD test set, which contains 2032 images.

\section{Experiments} \label{sec:exper}

To provide more comprehensive exploration of some detailed designs of our method, we conduct additional experiments, including Non-Prompt Generation, Inpainting With Different Mask, and more Ablation Study.

\subsection{Non-Prompt Generation}

\begin{figure*}
  \centering
  \setlength{\abovecaptionskip}{-0.05cm} 
  \includegraphics[width=1.0\textwidth]{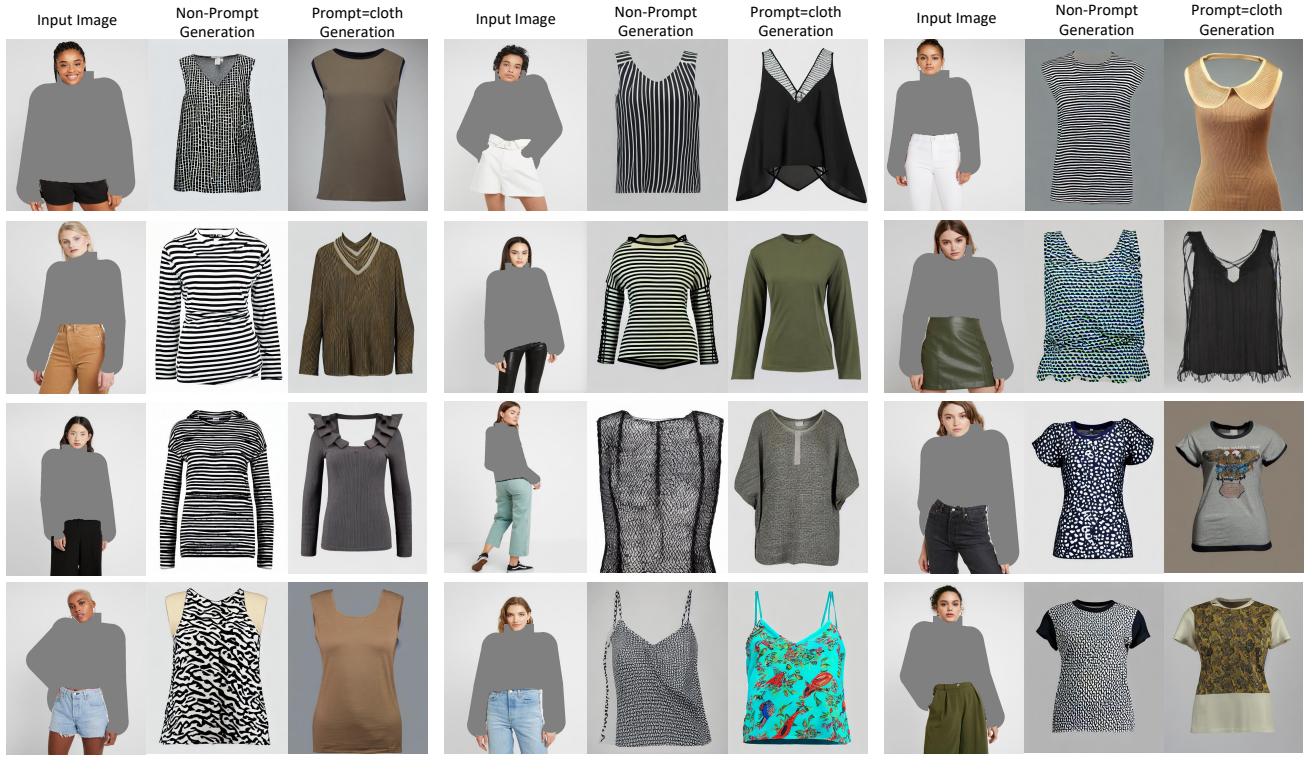}
  \caption{Comparison results with Non-Prompt Generation. In the absence of prompt, the model can still generate tops, but the clothing texture and color are monotonous. Therefore, we set the prompt to "cloth" to maintain the quality of generation while minimizing the influence of the prompt on recommendations.
  }
  \label{fig:noprompt}
  \vspace{-0.2cm}
\end{figure*}

To validate the impact of prompt choice to the generation performance, we implement the Non-Prompt model, where the Matching Clothes Diffusion Network (MCDN) takes empty prompt inputs, \textit{a.k.a.}, without any textual input. Specifically, when generating the matched clothes, the model randomly adds $T$ steps of Gaussian noise to a blank image as the initial image. The time steps $T$ and the prompt encoded representations are injected into the layers of the MCDN through the cross-attention mechanism. The object recognition of this network encoder can be guided by prompting words, while we set the prompting words to be empty. The generation model is only guided by the external clothing conditions, as shown in the Figure \ref{fig:noprompt}. Clearly, we can observe that without prompt words, the model is capable of generating images with clothing outlines under external shape constraints. However, the color and texture appear monotonous, lacking stylish features. Even though textual prompt plays a pivotal role in generating high-quality image, this work focuses on the matching-aware cloth generation. Therefore, we only use the simplest textual prompt of "cloth" to minimize the impact induced by complicated textual prompt. It is worthwhile and promising to study matching-aware generation models by integrating rich conditional information to the textual prompt, which will be explored in the future work.

\begin{figure*}[h]
  \centering 
  \setlength{\abovecaptionskip}{-0.05cm} 
  \includegraphics[width=1.0\textwidth]{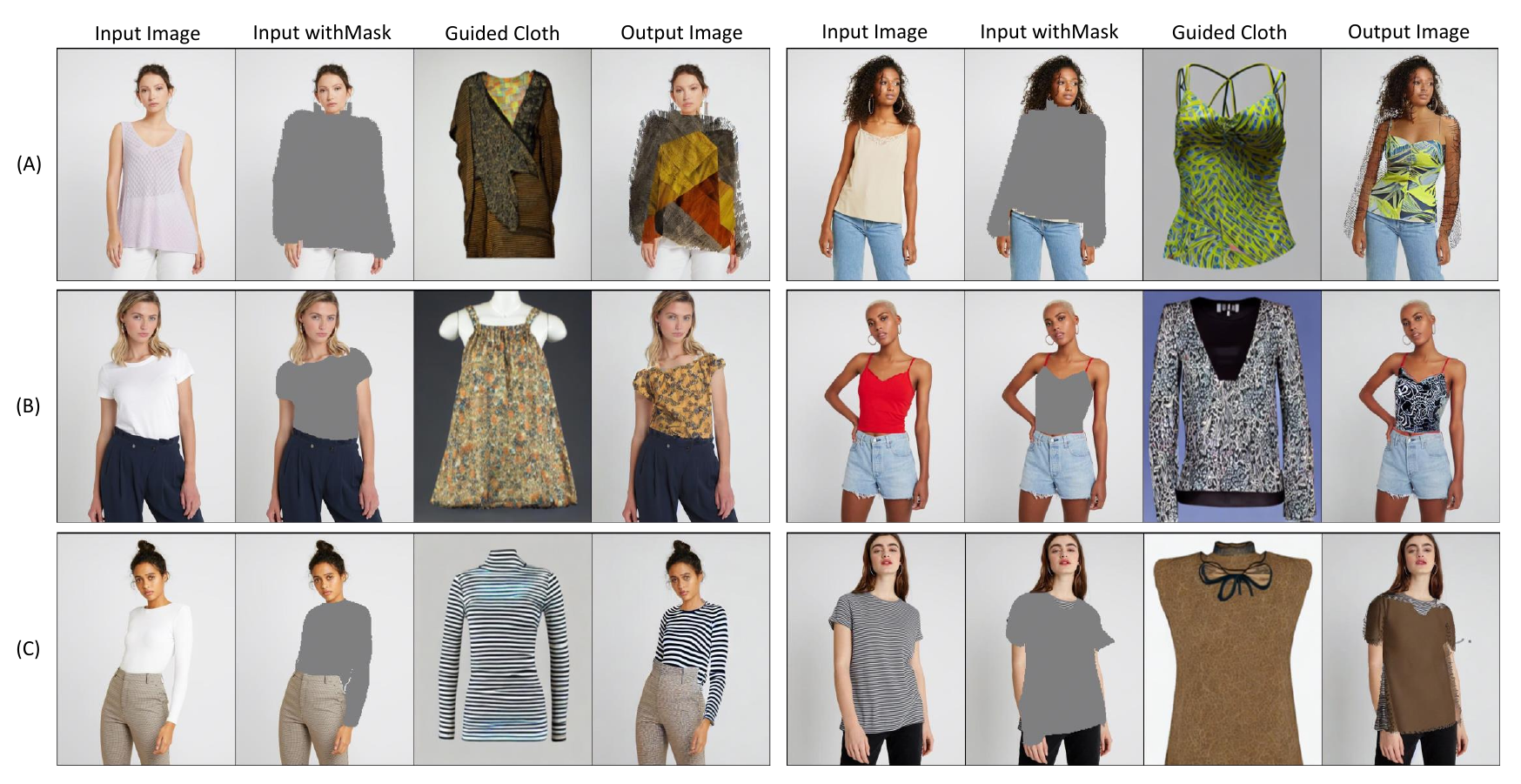}
  \caption{Inpainting results for different mask conditions. (A) employs a large mask that completely covers the clothing, (B) utilizes the mask of the original clothing worn on the body, and (C) adopts clothing mask warped by Try-on Generator.
  }
  \label{fig:denoise}
  \vspace{-0.3cm}
\end{figure*}

\begin{figure*}[h]
  \centering
  \includegraphics[width=1.0\textwidth]{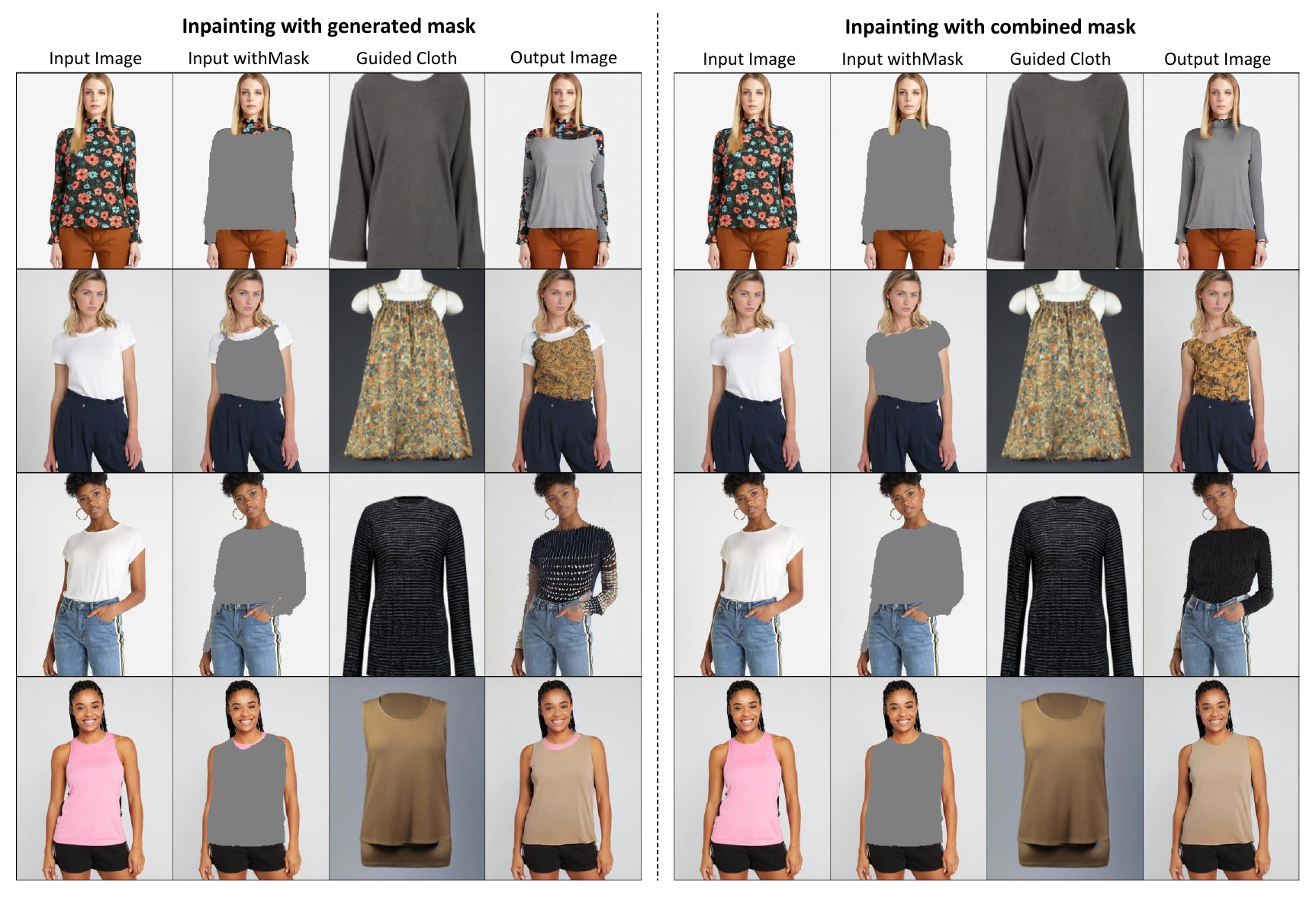}
  \caption{Inpainting results for different mask combinations. The left section employ masks warped by Try-on Condition Generator, while the right section uses a combined mask. It shows that the results on the right side do not retain any aspects of the original clothing.
  }
  \label{fig:suppl_mask}
\end{figure*}

\subsection{Inpainting With Different Masks}

In the Virtual Try-on Module, our framework inpaints the generated matching cloth to the masked region. The specific implementation of mask plays a crucial role for the final generation performance. As shown in Figure \ref{fig:denoise} (A), when the mask area is too large, the model alters the texture and shape of the Guided Cloth, extending the clothing to the entire mask range. When using the mask from the original worn clothing, as illustrated in Figure  \ref{fig:denoise} (B), the generated results can align well with the human body, while the clothing shape is influenced by the previous clothes. From Figure \ref{fig:denoise} (C) and the left side of Figure \ref{fig:suppl_mask}, we can observe that when the adaptive mask fails to cover the entire upper-body, the inpainting results of the final try-on image would include some noisy areas inherited from the original image.

To address the mask issue, we combine the original clothing mask with the adaptively generated mask from the Try-on Condition Generator. This combination ensures that the mask can cover the original image information while align with the distinctive shape of the Guided Cloth. The results, as shown on the right side of Figure \ref{fig:suppl_mask}, indicate that this approach effectively resolves the problem of adaptive clothing coverage. However, it comes with a weakness: the generation is highly dependent on the quality of the mask. To address this limitation, we employ the fine-tuned Paint-by-Example module to denoise the images.

\subsection{Ablation Study}

In this section, we present additional results and analysis in terms of two specifications, \ie the Shape Constraint and Matching Source.

\begin{figure}
  \centering
  \includegraphics[width=0.5\textwidth]{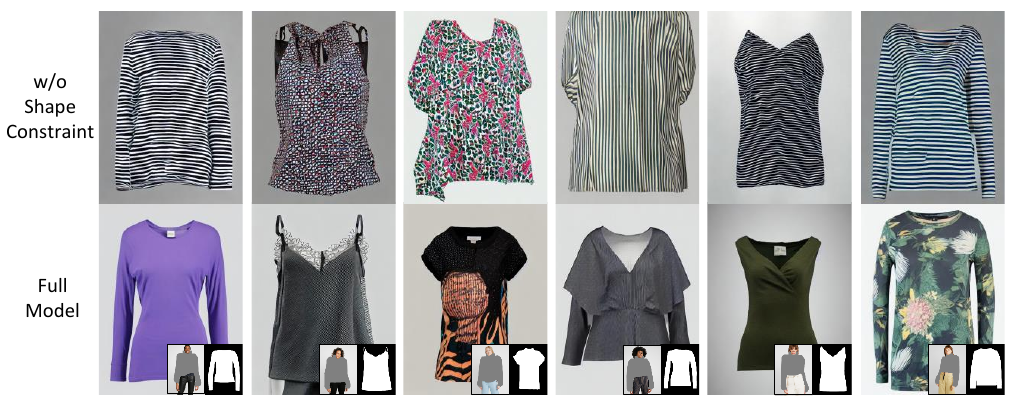}
  \caption{Results without Shape Constraint. The highlighted regions in the first row indicate that, without Shape Constraint, the generated clothes would miss some details or introduce redundant parts.
  }
  \label{fig:ab_shape}
\end{figure}

\textbf{w/o Shape Constraint:} In order to validate the necessity of the Shape Constraint Network, we disable it and present the generated clothes in the first row of Figure \ref{fig:ab_shape}. Obviously, even though the network is still to roughly capture the styles during generation, it is limited in preserving high-quality fine-grained details since the generated clothing may exhibit defects or protrusions. Furthermore, the boundary of different parts are unclear, resulting in blended sleeves and hemlines. In contrast, by applying clothing constraints, our model can generate clothes with rich texture details, as well as sharp and clear shape. These results demonstrate that the external guidance offered by the Shape Constraint is salient in generating high-quality clothes.

\begin{figure}
  \centering
  \includegraphics[width=0.5\textwidth]{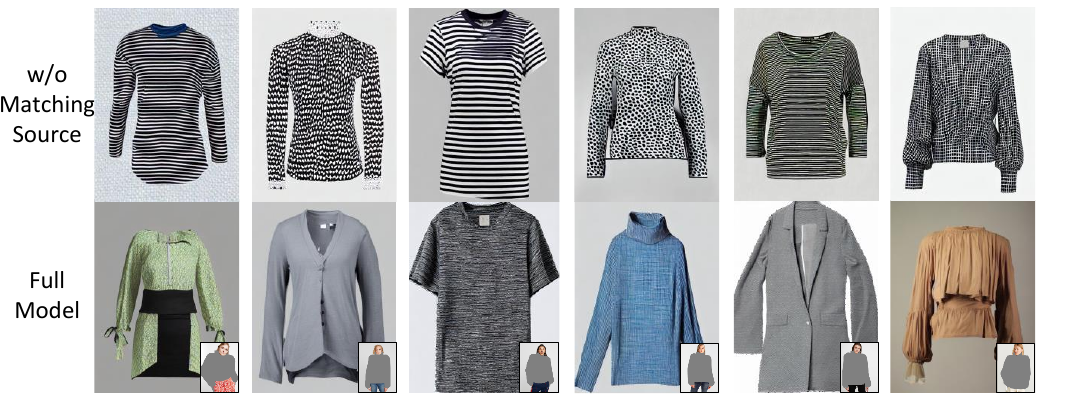}
  \caption{Comparison of generated images between with and without Matching Source. In the first row, the generated clothes exhibit monotonous texture and color, due to the absence of matching source information. This illustrates that our model can learn matching knowledge and recommend clothing suitable for the current attire of the user.
  }
  \label{fig:ab_ms}
\end{figure}

\textbf{w/o Matching Source:} 
We conduct an experiment to further demonstrate the utility of input source image during mix-and-match. Specifically, we remove the cloth from the given image and only keep the masked shape as the query, then we ask the model to generate a piece of cloth. We refer this setting as \textit{w/o Matching Source}, where the cloth in the given image is the source image for the mix-and-match task.
%the external guidance condition during clothing generation relies solely on the shape constraint. 
The results are shown in the first row of Figure \ref{fig:ab_ms}. We can observe that, without the matching source information, the generated clothes only adhere to the shape constraint but lack diversity in texture and color. In contrast, the clothes generated given the matching source clothes, as shown in the second row of Figure~\ref{fig:ab_ms}, are more diverse in color and texture. This observation implies that the generated clothes are truly conditioned on the given clothes in the matching source images, instead of unaware of the given clothes and falling into some indiscriminative common results.
%can learn matching knowledge and generate matching clothes with rich textures and coordinated colors for the input original image.

\section{User Study Details} \label{sec:userstudy}

We collaborate with fashion design researchers and design a professional protocol to evaluate the quality of mix-and-match. 
%To better assess the practical utility of the matching model, we invited fashion designers to provide a comprehensive evaluation of the matching results. 
The main reason of including fashion experts, especially researchers majoring in fashion design, lies in two perspectives. (1) Evaluating the fashionality of a pair of clothes is highly subjective and biased for laymen who have never systematically studied it. Meanwhile, fashion experts have been well trained, thus they can faithfully follow the given guidelines and give relatively more objective and professional evaluations. (2) Researchers in fashion design know how to curate a protocol to evaluate a research-oriented study, instead of judging an industry-level implementation. Therefore, the evaluation protocol could keep at a proper level, which is neither too rigorous, such as evaluating a designer's artworks, nor too simple. 
Following such motivation, our fashion domain experts design the evaluation protocol, which mainly includes two aspects: the rationality and diversity of clothing matching effects.

The evaluation of clothes matching effects typically focuses on three fashion elements, \ie \textit{style}, \textit{color}, and \textit{fabric}. First, \textit{style} means the proposed matching pair should adhere to the common pairing rules, such as top + bottom, top + jumpsuit, top + bottom + outerwear, \etc, and must avoid unconventional pairings, such as dress + bottom. Second, \textit{color} refers to the appropriate usage and combination of hues, including contrasting, monochromatic, and achromatic pairings. Finally, \textit{fabric} emphasizes that all the items in a matching set should be aligned to the same season in terms of the used fabric.

The diversity metric in clothing matching highlights that a good mix-and-match model should be able to generate diverse matching pairs, instead of similar pairs with limited variants. Diversity is significant in fashion design since fashion taste is highly personalized, and diverse options could meet more people's tastes. To be noted, the metric of diversity can not independently work, meanwhile, it should be under the constraint of the above mentioned metrics of style, color, and fabric.

In the fuzzy comprehensive evaluation method, the weights reflect the importance of each criterion in the evaluation system, making it an extremely crucial factor in the assessment process. Therefore, whether the weights are scientifically and reasonably determined directly impacts the accuracy of the evaluation. Hence, we adopt the expert estimation method, combining the automatically generated matching characteristics to determine the weight distribution for four evaluation criteria: style, color, fabric, and diversity. We invited five fashion experts who are knowledgeable about automatic generation technology to score four indicators on a five-level scale, as shown in the table \ref{tab:weight}.

\begin{table}
  \centering 
  \caption{Expert Evaluation Indicator Weight Scoring}
  \scalebox{0.75}{
  \begin{tabular}{c|ccccccc}
  \toprule Scale & Expert 1 & Expert 2 & Expert 3 & Expert 4 & Expert 5 & Total & Weight \\
  \midrule D1-Style & 4 & 2 & 4 & 3 & 4 & 17 & $24 \% $ \\
  D2-Color & 4 & 3 & 4 & 3 & 4 & 18 & $25 \% $ \\
  D3-Fabric & 3 & 2 & 4 & 3 & 3 & 15 & $21 \% $ \\
  D4-Variety & 4 & 4 & 4 & 4 & 5 & 21 & $30 \% $ \\

  \bottomrule
  \end{tabular}
}

  \label{tab:weight}
\end{table}

\section{Additional Results} \label{sec:results}

\begin{figure*}
  \centering
  \includegraphics[width=1.0\textwidth]{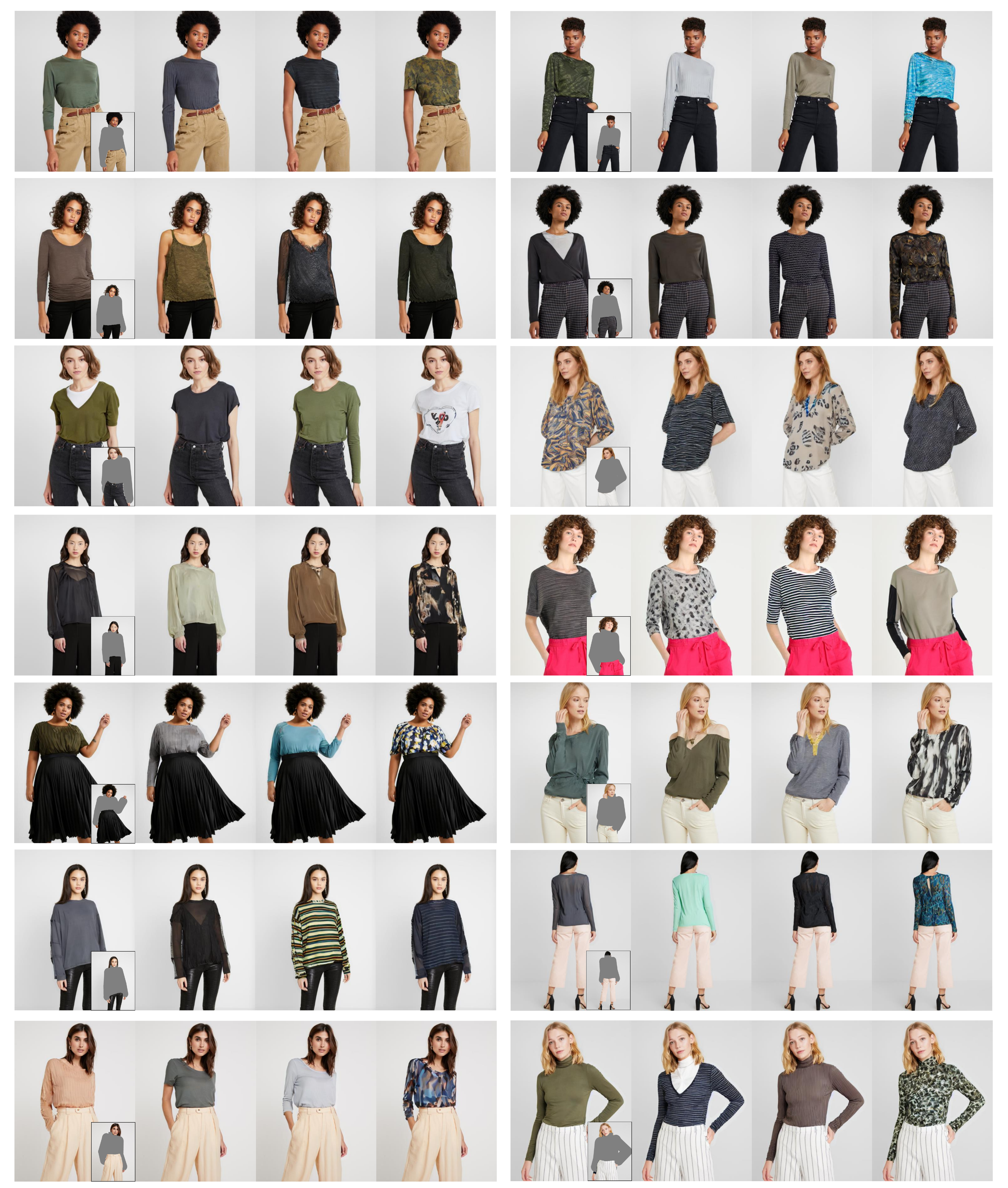}
  \caption{More results of the generated images. It can be observed that our proposed one-stop Smart Fitting Room system can generate images with diverse styles and improved matching combinations.
  }
  \label{fig:suppl_result}
\end{figure*}

In order to demonstrate our model's capability of mix-and-match, we present more examples generated by our method, as shown in Figure \ref{fig:suppl_result}. We would like to emphasize that the clothes generated by our model are well matched to the given clothes and tailored to the styles of certain people. Specifically, in terms of clothing generation, our model's generation results exhibit rich texture details, featuring soft and skin-friendly fabrics, as well as thin and sheer materials. Regarding matching combinations, our results demonstrate a diverse and harmonious color palette, presenting various styles, such as short sleeves, mid-sleeves, and long sleeves. Furthermore, in terms of the try-on effect, the final output images present realistic adherence to the human body contour and high-quality texture.

\newpage

\end{document}